\documentclass[twocolumn]{iopart}
\usepackage{graphicx,color}
\begin{document}

\title[Probing Wigner correlations in a suspended carbon nanotube]{Probing Wigner correlations in a suspended carbon nanotube}

\author{N. Traverso {Ziani}$^{1,2}$, F. {Cavaliere}$^{1,2}$, M. {Sassetti}$^{1,2}$}

\address{Dipartimento di Fisica, Universit\`a di Genova, Via Dodecaneso 33,
  16146, Genova, Italy.}
\address{CNR-SPIN, Via Dodecaneso 33,
  16146, Genova, Italy.}

\ead{cavalier@fisica.unige.it}
\begin{abstract}
The influence of the electron-vibron coupling on the transport properties of a strongly interacting quantum dot built in a suspended carbon nanotube is analyzed. The latter is probed by a charged AFM tip scanned along {the} axis {of the CNT} which induces oscillations of the chemical potential and of the linear conductance. These oscillations are due to the competition between finite-size effects and the formation of a Wigner molecule for strong interactions. Such oscillations are shown to be suppressed by the electron-vibron coupling. The suppression is more pronounced in the regime of {weak} Coulomb interactions, which ensures that probing Wigner correlations in such a system is in principle possible.
\end{abstract}

\maketitle
When Coulomb interaction dominates over the kinetic energy of electronic systems, Wigner molecules of electrons can emerge~\cite{vignale}. After more than 75 years~\cite{wigner}, however the details of the appearance of such a strongly correlated state of matter are only partially known: the theoretical modeling and the experimental realization of Wigner molecules are still challenging problems~\cite{wigmol1,wigmol2,1Dwig}, even though quantum dots~\cite{koudots}, allowed a breakthrough in the problem. As long as the theoretical modeling is concerned, the description of 2-dimensional (2D) Wigner molecules has relied mainly on numerical techniques~\cite{2Dnum2,Egger99,2Dnum3,2Dnum5,2Dnum6,2Dnum7,serra,2Dnum8,2Dnum10}, while in 1 dimension (1D), aside from numerical techniques~\cite{kramer,xia,bortz,pederiva,bedu,szafran,wire3,polini,shulenburger,secchi1,sgm2,astrak,burke,polini2,silva}, analytical methods can be employed~\cite{giamarchi,schulz,mantelli,safi,sablikov,fiete1,075}, mainly thanks to the Luttinger liquid theory~\cite{giamarchi,haldanefluid,voit,delft}. On the experimental side, in 2D optical spectroscopy has been employed~\cite{wigmol1,wigmol2}; in 1D transport techniques based on momentum resolved tunneling between quantum wires~\cite{fili2,wire1,wire2}, and on the magnetic properties of the Wigner molecule~\cite{nature} have been used. Recently other experimental set ups for the detection of 1D Wigner molecules have been proposed~\cite{sgm2,AFM,secchi2,sgm1,linear,epl,njp}: the transport properties of 1D quantum dots perturbed by local probes such as AFM and STM tips have been demonstrated to be effective in the detection of the Wigner molecule. {In} many of the experimental realizations{,} quantum dots are however surrounded by metallic gates which unavoidably screen the Coulomb interactions, demoting the formation of the molecule itself. A possible candidate in which screening effects can be dramatically reduced is a suspended carbon nanotube (CNT)~\cite{nature,secchiarxiv}, a system in which rather strong Coulomb interactions can be attained. In such a system, however, the electronic degrees of freedom are strongly coupled to the vibrational ones as it occurs in nano-electro-mechanical systems (NEMS)~\cite{braig,nems1,nems2,4,10,15,6}: is has indeed been demonstrated that suspended CNTs can behave as mechanical resonators~\cite{resonator}. Such a source of fluctuations can potentially be detrimental to the formation of the Wigner molecule.\\
In this work we investigate the transport properties of a one dimensional quantum dot, built in a suspended and interacting CNT, scanned by a negatively charged AFM tip free to move along the axis of the CNT. Employing the Luttinger liquid theory, we address both the renormalization induced on the chemical potential and on the linear conductance peak by  the presence of the AFM tip. In the absence of electron vibron coupling these quantities have been demonstrated to be effective in the detection of the Wigner molecule. {We} show that care must be taken, since the coupling to a vibron indeed suppresses the oscillations induced by the tip. However, since electron-electron interaction reduces the effect of the electron-vibron coupling, we can conclude that a strongly interacting, suspended CNT may be an excellent candidate to investigate the physics of Wigner molecules.\\
As a first step we briefly remind the low energy model for finite size interacting CNTs. It is well established~\cite{egger} that the low energy properties of any finite interacting metallic (or semiconducting, if the Fermi point is not close to the charge neutrality point~\cite{noi}) CNT can be described in terms of a four-channel Luttinger liquid. The Hamiltonian ($\hbar=1$) ${H}$ is ${H}=\sum_{j=1}^4{H}_{j}$, with~\cite{yo,grifoni}
\begin{equation}
\label{eq:hambos}
{H}_{j}=\frac{1}{2}E_{j}\left({N}_{j}-\delta_{1,j} N_g\right)^{2}+\sum_{\mu\geq 1}\mu\omega_{j}{b}_{j,\mu}^{\dagger}{b}_{j,\mu}\, ,
\end{equation}
where $E_j=\pi v_F/(4Lg_j^2)$, $v_F$ is the Fermi velocity and $g_j$ is the Luttinger liquid parameters of each channel, with $0<g_1\equiv g\leq 1$ and $g_2=g_3=g_4=1$. For a metallic CNT $v_{F}=8\cdot 10^5\ \mathrm{m/s}$~\cite{saito}, while considerably lower velocities can be found in semiconducting CNTs~\cite{noi}. Here,  ${N}_j$ is the number of excess electrons in the channel $j$ with $N_1$ the total number of excess electrons, and $N_g\propto V_g$ is due to the inclusion of a gate contact capacitively coupled to the dot. Finally, the second term to the r.h.s. of Eq.~(\ref{eq:hambos}) describes collective excitations with ${b}_{j,\mu}$ bosonic operators and $\omega_j=\pi v_F/Lg_j$. The electron field operator reads
\begin{equation}
\label{eq:fieldopop}
{\Psi}_{s}(\mathbf{r})=\sum_{r=\pm 1}\sum_{\alpha=\pm 1}f_{r,\alpha}(\mathbf{r})e^{irq_{\mathrm{F}}x}{\psi}_{+1,r\alpha,s}(rx),
\end{equation}
where
\begin{eqnarray}
{\psi}_{+1,\alpha,s}(x)&=&\frac{{\eta}_{\alpha,s}}{\sqrt{2\pi\tilde{a}}}e^{-i\theta_{\alpha,s}}e^{i\frac{\pi
    x}{4L}\left({N}_{1}+\alpha
  {N}_{2}+s{N}_{3}+\alpha s
  {N}_{4}\right)}\cdot\nonumber\\ &&e^{\frac{i}{2}\left[{\phi}_{1}(x)+\alpha{\phi}_{2}(x)+s{\phi}_{3}(x)+\alpha
    s {\phi}_{4}(x)\right]}\, ,\label{eq:fieldopdiag}
\end{eqnarray}
and
\begin{eqnarray}
\phi_{j}(x)&=&\sum_{\mu\geq 1}\left\{\frac{\cos{(q_{\mu}x)}}{\sqrt{\mu g_{j}}}\left[{b}_{j,\mu}+{b}_{j,\mu}^{\dagger}\right]\nonumber\right.\\
&+&\left.i\sqrt{\frac{g_{j}}{\mu}}\sin(q_{\mu}x)\left[{b}_{j,\mu}-{b}_{j,\mu}^{\dagger}\right]\right\}\, .
\end{eqnarray}
One has $q_{\mu}=\pi\mu/L$, $[{\theta}_{\alpha,s},{N}_{\alpha',s'}]=i\delta_{s,s'}\delta_{\alpha,\alpha'}$, $N_{\alpha,s}={N}_{1}+\alpha
  {N}_{2}+s{N}_{3}+\alpha s
  {N}_{4}$ and $\tilde a\ll L$ a cutoff length.
The functions  $f_{r,\alpha}(\mathbf{r})$ consist of a superposition of wavefunctions for $p_{z}$
orbitals, peaked around the positions of atoms in the CNT~\cite{saito}. Finally, $q_F$ is the distance in the reciprocal lattice between a Fermi point and the closest Dirac point, considering an effective n doping for the CNT~\cite{noi}. Throughout this work, we will assume a reference state with $N_{0}=4\kappa$ (with $\kappa>0$ an integer) electrons, and $q_{F}=\pi N_{0}/4L$.\\
\noindent When the CNT is suspended the interplay between the electronic and the mechanical degrees of freedom must be addressed. Among the possible quantized oscillation modes (vibrons)~\cite{suzu,mart,penn,mahan,eros,flens,alves,izu}, the most relevant for the electronic properties considered in this work are the lowest stretching ones~\cite{sapmaz}. We assume that the vibron extends over the whole length of the CNT, clamped at the CNT ends. The vibron Hamiltonian is
\begin{equation}
\label{eq:vibh0}
{H}_{\mathrm v}=\frac{{P_{0}}^2}{2M}+\frac{M\omega_{0}^2}{2}{X}_{0}^2\, ,
\end{equation}
where $\omega_{0}=\pi v_{\mathrm s}/{L}$
with  $v_{\mathrm s}\approx 2.4\cdot 10^{4}$ m/s and
$M=2\pi \mathcal{W}{L}\rho_{0}$ is the CNT mass, with $\mathcal{W}$ the radius and
$\rho_{0}\approx6.7\cdot10^{-7}$ Kg/$\mathrm{m}^2$. Here, $\sqrt{2M}X_{0}=\left(b_{0}+b_{0}^{\dagger}\right)$ is the amplitude operator of the strain field in the fundamental mode ${u}(\mathbf{r})=\sqrt{2}{X_{0}}\sin\left(\frac{\pi x}{{L}}\right)$.
In the elastic limit, the form of the Hamiltonian ${H}_{\mathrm{d-v}}$ coupling electronic and vibronic degrees of freedom is that of a deformation potential $H_{\mathrm{d-v}}=c\int\ \mathrm{d}x\ R(\mathbf{r})\partial_{x}u(x)$, where $R(\mathbf{r})=\sum_{s}\Psi^{\dagger}_{s}(\mathbf{r})\Psi_{s}(\mathbf{r})$ is the electron density and $c\approx 30\ \mathrm{eV}$.\\
Neglecting rapidly oscillating terms with a wavelength $\approx a$ (with $a$ the CNT lattice constant), the electron vibron coupling simplifies as~\cite{egger,noi,proc} $H_{\mathrm{d-v}}\sim c\int_{0}^{L}\ \mathrm{d}x\ \rho(x)\partial_{x}u(x)$.  Keeping only the lowest relevant harmonics in the density bosonic expansion~\cite{sablikov,haldanefluid} one obtains ${\rho}(x)=\rho^{LW}(x)+F\rho^F(x)+(1-F)\rho^W(x)$ where $\rho^{LW}(x)$ is the long wave contribution, $\rho^F(x)$ are the Friedel oscillations~\cite{yo}, due to finite size effects, and $\rho^W(x)$ is the Wigner contribution, due to interaction effects. The parameter $0\leq F\leq 1$ cannot be calculated within the Luttinger liquid theory: numerically one finds $F=1$ in the non interacting case, while $F\ll 1$ in the limit of very strong interactions~\cite{bortz}. In the bosonization language one has
\begin{eqnarray}
\rho^{LW}(x)\!&=&\!\frac{N_1}{L}+\frac{4q_F}{\pi}-\frac{\partial_x}{\pi}\sum_{\alpha,s}\varphi_{\alpha,s}(x)\nonumber\\
\rho^F(x)\!&=&\!-\sum_{\alpha,s}\frac{N_{\alpha,s}}{L}\cos\left[\!\mathcal{L}_{F}(x)\!+\!\frac{2\pi x N_{\alpha,s}}{L}\!-\!2\varphi_{\alpha,s}(x)\!\right]\!,\nonumber\\
\rho^W(x)\!&=&\!-\frac{N_1}{L}\cos\left[\mathcal{L}_{W}(x)+\frac{2\pi x N_{1}}{L}-4\varphi_{1}(x)\right],\nonumber
\end{eqnarray}
with $\mathcal{L}_{F}(x)=2q_{F}(x)-2g^2h(x)$, $\mathcal{L}_{W}(x)=4\mathcal{L}_{F}(x)$, $h(x)=[\phi_1(x),\phi_1(-x)]/(4i)$, and $\varphi_{\alpha,s}(x)=[\phi_{\alpha,s}(-x)-\phi_{\alpha,s}(x)]/2$, $\varphi_{1}(x)=[\phi_{1}(-x)-\phi_{1}(x)]/2$.\\
The long-wave part has a typical length scale $\approx L$, while the Friedel and Wigner contributions oscillate with typical wavelengths $(2q_{F})^{-1}$ and $(8q_{F})^{-1}$ respectively. In this work, we will consider a CNT dot with with a not too small number of electrons $N\gg 1$ so that the Friedel and Wigner terms of the density oscillate much faster than the strain field. As a result, the only relevant contribution to $H_{\mathrm{d-v}}$ stems from the long-wave part of the density and reads
\begin{equation}
\label{eq:ephint00}
{H}_{\mathrm{d-v}}=\frac{c}{2\pi}\int_{0}^{L}\mathrm{d}x\ \partial_{x}\left[\phi_{1}(x)-\phi_{1}(-x)\right] \partial_{x}{u}_{p}(x)\, .
\end{equation}
Introducing $B_{\mu}=ib_{1,\mu}$ and $\sqrt{2\mu\omega_{1}}X_{\mu}$ (with a conjugated $P_{\mu}$ satisfying $[X_{\mu},P_{\nu}]=\delta_{\mu,\nu}$) the electron-vibron coupling casts into
\begin{equation}
\label{eq:ephint00}
{H}_{\mathrm{d-v}}=\sqrt{M}X_{0}\sum_{\mu\geq 1}C_{\mu}X_{\mu}\, .
\end{equation}
where $\quad C_{\mu\geq 1}=2\lambda_{\mathrm{m}}\omega_{0}^{3/2}\sqrt{\omega_{1}}L^{-1}\int_{0}^{L}{\mathrm d}x\ \cos\left(\frac{\mu\pi x}{L}\right)\cos\left(\frac{\pi}{{L}}x\right)$, with $c\lambda_{\mathrm m}^{-1}=\sqrt{\rho_{0}\pi\mathcal{W}v_{\mathrm s}}$. For a typical CNT, $\lambda_{\mathrm{m}}\approx 2$~\cite{noi}. We will assume this value in the following.\\
It can be readily seen that the operator ${h}_1={H}_1+{H}_{\mathrm{d-v}}+H_{V}$ is manifestly quadratic in the operators $X_{0}$, $P_{0}$, $X_{\mu}$, and $P_{\mu}$ and thus can be diagonalized exactly~\cite{ullersma}. The diagonal form of $h_{1}$ is
\begin{equation}
{h}_1=\frac{1}{2}E_{1}\left({N}_{1}-N_{g}\right)^{2}+\sum_{\mu\geq0}\left(\frac{{\bar{P}}_{\mu}^{2}}{2}+\Omega_{\mu}^{2}\frac{{\bar{X}}_{\mu}^{2}}{2}\right)\, .
\end{equation}
For the case of a vibron in its fundamental mode, spanning the entire dot, one finds $C_{\mu}=\lambda_{m}\omega_{0}^{3/2}\sqrt{\omega_{1}}\delta_{\mu,1}$, i.e. the vibron couples to the lowest-lying plasmon mode only. As a result, $\Omega_{\mu\geq 2}\equiv\mu\omega_{1}$ with the corresponding plasmonic modes remaining completely unaffected. The energy of the two lowest-lying collective modes are the positive roots of $\left(\epsilon^2-1\right)\left(\epsilon^2-r^2\right)=\lambda_{\mathrm{m}}^{2}r$ where $\epsilon=\Omega_{\mu}/\omega_{0}$ ($\mu=0,1$) and $r=\omega_{1}/\omega_{0}=v_{F}/(g v_{s})$. Clearly, the parameter $r$ governs the nature of the solutions. Even in semiconducting CNTs, one finds $r>1$ since $v_{s}$ does not exceed $v_{F}$. In this regime, it can be readily seen that the mode with $\mu=0$ has energy $\Omega_{0}<\omega_{0}$ and represents a vibron dressed by the plasmonic mode, while $\Omega_{1}>\omega_{1}$ with $\Omega_{1}\approx\omega_{1}$ is the energy of a collective mode almost unaffected by the coupling. As a consequence, we can further simplify $h_{1}$ neglecting the slight renormalization of the first plasmonic mode, to obtain
\begin{equation}
\label{eq:hamcolfin}
{h}_1=H_{1}+\frac{{\bar{P}}_{0}^{2}}{2}+\Omega_{0}^{2}\frac{{\bar{X}}_{0}^{2}}{2}\, .
\end{equation}
Within the above approximation, the operator $\phi_{1}(x)$, entering the field operators $\Psi_{s}(x)$ transforms upon the diagonalization into $\bar{\phi}_{1}(x)=\phi_{1}(x)+\phi_{0}(x)$, with the contribution
\begin{equation}
\label{eq:phi0}
\phi_{0}(x)=\alpha_{0}(x)\bar{X}_{0}+\beta_{0}(x)\bar{P}_{0}\, ,
\end{equation}
stemming from the coupled vibronic mode. We find
\begin{equation}
\label{eq:alpha0}
\alpha_{0}(x)\approx-\sqrt{2g\omega_{1}}\lambda_{\mathrm{m}}r^{-3/2}k_{0}\sin{\left(\frac{\pi x}{L}\right)}
\end{equation}
and $\beta_{0}(x)=\left[L/(\pi g\omega_{1})\right]\partial_{x}\alpha_{0}(x)$ with $k_{0}^{-2}=1-\lambda_{m}^{2}/r^{2}$. \\
The total Hamiltonian $H_d$ of the vibrating CNT  $H_d=h_1+H_2+H_3+H_4$ is separated in two additive contributions: the one of zero modes $H_N$, which only depends on $N_{i}$, $i=1,..,4$ and the bosonic one $H_b$. The eigenstates are $|\{N_i\},\{n^{(\mu)}_{i}\}\rangle$, where  $n^{(\mu)}_i$ is the number of bosonic excitations in the $i$-th channel, with momentum $\pi\mu/L$. For a given number of particles $N$, the ground state $|N\rangle$ is obtained minimizing the zero mode energy with the constraint $N_1$ and setting $n^{(\mu)}_{i}=0$ $\forall\mu,i$.\\
We now turn to the transport properties of the suspended CNT, tunnel coupled to two lateral contacts and capacitively coupled to a charged AFM tip, located at $0\leq x_0\leq L$ and kept at a potential $V_{tip}<0$. A scheme of the set up is shown in Fig.1.
\begin{figure}
\begin{center}
  \includegraphics[width=6.8cm,keepaspectratio]{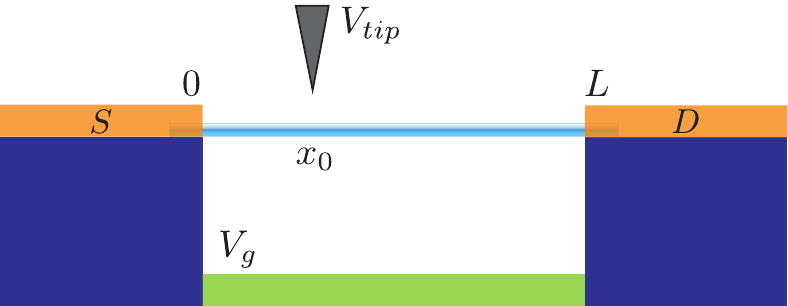}
  \caption{Scheme of the setup. See text for further details.}
  \label{fig:fig1}
  \end{center}
\end{figure}
The source and drain ($\lambda=S,D$) contacts are non-interacting Fermi gases at potential $-V/2$ and $V/2$ respectively, with Hamiltonians $H_\lambda$. The tunneling Hamiltonians connecting the dot and the lead $\lambda$ read~\cite{noi,proc,bercioux,giacomo}
\begin{equation}
{H}_{\lambda}^{t}=t_{0}\sum_{\lambda=S,D}\sum_{\alpha,s,q}{\psi}_{+1,\alpha,s}(x_{\lambda}){c}_{\lambda,s}(q)+\mathrm{h.c.}\,,
\end{equation}
where we assumed symmetric barriers with $t_{0}$ the tunneling amplitude. Here, ${c}_{\lambda,s}(q)$ are the operators for an electron with momentum $q$ and spin $s$ in the non-interacting lead $\lambda$ and ${x}_{1}=0$, ${x}_{2}=L$ are the
positions of the tunneling contacts.\\
The capacitive coupling between the dot and the tip (assumed non magnetic~\cite{biggio}) is given by $H_{AFM}=H_F+H_W$ with,
\begin{eqnarray}
H_F&=&V^{(F)}\sum_{\alpha,s}\cos\left[\mathcal{L}_{F}(x)+\frac{2\pi N_{\alpha,s} x}{L}-2{\varphi}_{\alpha,s}(x)\right],\nonumber\\
H_W&=&V^{(W)}\cos\left[\mathcal{L}_{W}(x)+\frac{2\pi N_1 x}{L}-4{\varphi}_{1}(x)\right].\nonumber \end{eqnarray}
Here, $V^{(\xi)}\propto|V_{tip}|$ ($\xi=F,W$) parameterizes the strength of coupling between the tip and the Friedel or Wigner contributions to the density. We have assumed in this work a sufficiently sharp tip, whose width $\delta$ is larger than the CNT lattice constant but smaller than the wavelength of the Wigner oscillations. The coupled tip induces a renormalization of both the chemical potential of the dot and of its linear conductance.\\
To the lowest perturbative order in $V^{\xi}$ is given by $\mu=\mu_{0}+\delta\mu(x_{0})$ where $\mu_{0}=\langle N+1|H_{d}|N+1\rangle-\langle N|H_{d}|N\rangle$ is the chemical potential in the absence of  the tip and $\delta\mu(x_0)=\langle N+1|H_{AFM}|N+1\rangle-\langle N|H_{AFM}|N\rangle$ the first order tip correction. Here, and in the rest of the paper, for definiteness we assume $N_1=4n$ (with $n\geq 0$ an integer). One finds
\begin{equation}
\delta\mu(x_0)=\sum_{\xi=F,W}V^{(\xi)}[\zeta^{(\xi)}(N+1,x_0)-\zeta^{(\xi)}(N,x_0)]\, ,
\end{equation}
where
\begin{eqnarray}
\zeta^{(F)}(n,x)\!&=&\!K(x)^{\frac{3+g}{4}}V(x)^{\frac{1}{4}}\!\sum_{\alpha,s}\!\cos\left[\mathcal{L}_F(x)\!+\!\frac{2\pi x N_{\alpha,s}}{L}\right],\nonumber\\
\zeta^{(W)}(n,x)\!&=&\!K(x)^{4g}V(x)^{4}\cos\left[\mathcal{L}_W(x)\!+\!\frac{2\pi x n}{L}\right],\label{eq:basta}\\
K(x)\!&=&\!\frac{\sinh\left(\frac{\pi\tilde{a}}{2L}\right)}{\sqrt{\sinh^2\left(\!\frac{\pi\tilde{a}}{2L}\right)\!+\!\sin^2\left(\frac{\pi x}{L}\!\right)}}\ \ ;\,\,V(x)\!=\!e^{-\frac{\alpha^2_0(x)}{\Omega_0}}.\nonumber
\end{eqnarray}
In Eq.~(\ref{eq:basta}) we have $N_{\alpha,s}=N_1/4$ if $n=N_1$, while $N_{\alpha^*,s^*}=N_1/4+1$ for a given $\alpha=\alpha^*$, $s=s^*$ and $N_{\alpha\neq\alpha^*,s\neq s^*}=N_1/4$ if $n=N_1+1$. This is due to the fourfold degeneracy of the ground state with $N+1$ electrons. The chemical potential corrections consist of an oscillating pattern, oscillating in accordance to the Friedel or the Wigner length scale, modulated by an envelope function slowly varying on the scale of $L$. The envelope functions are composed by a term $K(x)$, stemming from the collective modes of the CNT at energies $\mu\omega_{1}$ and present also in the absence of the electron-vibron coupling~\cite{AFM}, and by $V(x)$ which originates from the electron-vibron coupling. Both $K(x)\leq 1$ and $V(x)\leq 1$ contribute to suppress the oscillatory pattern: the coupling of mechanical and electrical degrees of freedom leads therefore to an {\em additional} suppression of both Friedel and Wigner oscillations, induced by increased fluctuations of the charge degree of freedom. The term $K(x)$ exhibits different power laws as a function of the Luttinger parameter $g$ in the Friedel and in the Wigner channels: the suppression of the Wigner fluctuations is most severe than that of the Friedel channel when $g\to 1$, while in the strong interactions regime one has no suppression of the Wigner oscillations due to the high energy collective modes of the CNT, in contrast with the Friedel oscillations which are still damped even for $g=0$.\\
In the sequential tunneling regime, the linear conductance can be evaluated setting up a rate equation for the occupation probability of the dot states. Tunneling rates between dot ground states with $N=N_{0}+N_{1}$ and $N+1$ electrons are evaluated to the second order in $t_{0}$ by means of the Keldysh technique~\cite{AFM,blum,haupt,master}. The effects of the AFM tip are evaluated as a perturbation to first order in $V^{(\xi)}$. In the linear regime and for low temperature $k_{B}T<\Omega_{0}$ tunneling rates attain the form $\Gamma^{(\lambda)}_{N\to N+1}=\gamma^{(\lambda)}(x_{0})f(\mu)$ and $\Gamma^{(\lambda)}_{N_1+1\to N_1}=\gamma^{(\lambda)}(x_{0})f(-\mu)$ where $f(E)=\left[1+e^{\beta E}\right]^{-1}$ is the Fermi function with $\beta^{-1}=k_{B}T$. The tunneling rates $\gamma^{(\lambda)}(x_{0})$ can be evaluated explicitly following a procedure analogous to that outlined in Ref.~\cite{AFM}. Here we just quote the final result $\gamma^{(\lambda)}(x_{0})=\gamma_{0}^{(\lambda)}\left[1+\delta\gamma^{(F)}(x_0)+\delta\gamma^{(W)}(x_{0})\right]$ with $\gamma_{0}^{(\lambda)}=\nu_{0}|t_{0}|^{2}(\pi\tilde{a})^{-1}\left(1-e^{-\pi \tilde{a}/L}\right)^{-(3+g)/4}$ where $\nu_{0}$ is the leads density of states, and
\begin{equation}
\delta\gamma^{(\xi)}= 2V^{(\xi)}N^{(\xi)}(x_{0})\sum_{\mathbf{m}\neq\mathbf{0},\delta={i,f}}\frac{1}{\Lambda}
B^{(\xi,\delta)}_{\mathbf{m}}C_{\mathbf{m}}^{(\delta,\xi)}(x_0)\, .\nonumber
\end{equation}
Several quantities have been introduced:  $\mathbf{m}=(m_1,m_2,m_3,m_4)$ is a vector of four integers,$\Lambda=\epsilon_\rho(m_1+m_2)+\epsilon_\sigma(m_3+m_4)$, the coefficients are given by $B^{F,i}_{\mathbf{m}}=b_{m_{1}}^{+,1/4}b_{m_{2}}^{-,1/4}b_{m_{3}}^{+,3/4}b_{m_{4}}^{-,3/4}$, and, $B^{W,i}_{\mathbf{n},\mathbf{m}}=b_{m_{1}}^{+,1}b_{m_{2}}^{-,1}\delta_{m_{3},0}\delta_{m_{4},0}$, while ${B}_{\mathbf{n},\mathbf{m}}^{\xi,f}$ is expressed in terms of $B_{\mathbf{m}}^{\xi}$ as
${B}_{m_{1},m_{2},m_{3},m_{4}}^{\xi,f}=B_{m_{2},m_{1},m_{4},m_{3}}^{\xi,i}$
with
\begin{eqnarray}
b_{l}^{+,\kappa}&=&\left(-e^{-\frac{\pi\alpha}{L}}\right)^{l}\left(1-e^{-\frac{\alpha\pi}{L}}\right)^{\kappa}\frac{\Gamma(1+\kappa)\theta(l)}{l!\Gamma(1+\kappa-l)}\\
b_{l}^{-,\kappa}&=&\left(e^{-\frac{\pi\alpha}{L}}\right)^{l}\left(1-e^{-\frac{\alpha\pi}{L}}\right)^{-\kappa}\frac{\Gamma(l+\kappa)}{l!\Gamma(\kappa)}\theta(l).
\end{eqnarray}
The oscillations of the tunneling rate as a function of the tip position are encoded in the functions $C_{\mathbf{n},\mathbf{m}}^{(\delta,\xi)}(x_0)$, given by
$$C_{\mathbf{m}}^{(\delta,\xi)}(x_0)\!=\!\cos\left[\!\mathcal{L}^{(\delta)}_{F/W}(x_0)\!+\!\frac{\pi(m_{1}\!+\!m_{3}\!-\!m_{2}\!-\!m_{4})x_0}{L}\!\right],$$
with $\mathcal{L}_{F}^{(i)}(x_0)=\mathcal{L}_{F}(x_{0})+\pi N_1 x_{0}/2L$, $\mathcal{L}_{W}^{(i)}(x_0)=\mathcal{L}_{F}(x_{0})+2\pi N_1 x_{0}/L$, $\mathcal{L}^{(f)}_{F}(x_0)=\mathcal{L}^{(i)}_{F}(x_0)+\frac{\pi x_0}{2L}$, $\mathcal{L}^{(f)}_{W}(x_0)=\mathcal{L}^{(i)}_{W}(x_0)+\frac{2\pi x_0}{L}$, and finally the pre-factors $N^{(\xi)}(x_{0})$ are given by
$N^{(F)}(x_0)=K(x_0)^{\frac{3+g}{4}}V(x_0)^{\frac{1}{4}}$ and $N^{(W)}(x_0)=K(x_{0})^{4g}V(x_0)^{4}$.\\
In terms of the above quantities, the differential conductance $G$ is given by~\cite{AFM}
\begin{eqnarray}
G&=&\frac{4\beta e^2 \Delta(x_{0})f(-\mu)}{4+e^{\beta\mu}}\label{eq:cond}\\
\Delta(x_{0})&=&\frac{\gamma^{(S)}(x_{0})\gamma^{(D)}(x_{0})}{\gamma^{(S)}(x_{0})+\gamma^{(D)}(x_{0})}\, .
\end{eqnarray}
The factor 4 in Eq.~(\ref{eq:cond}) stems from the fourfold degeneracy of the ground state with $N_1+1$ electrons. We now turn to a discussion of the results.
\begin{figure}
\begin{center}
  \includegraphics[width=7cm,keepaspectratio]{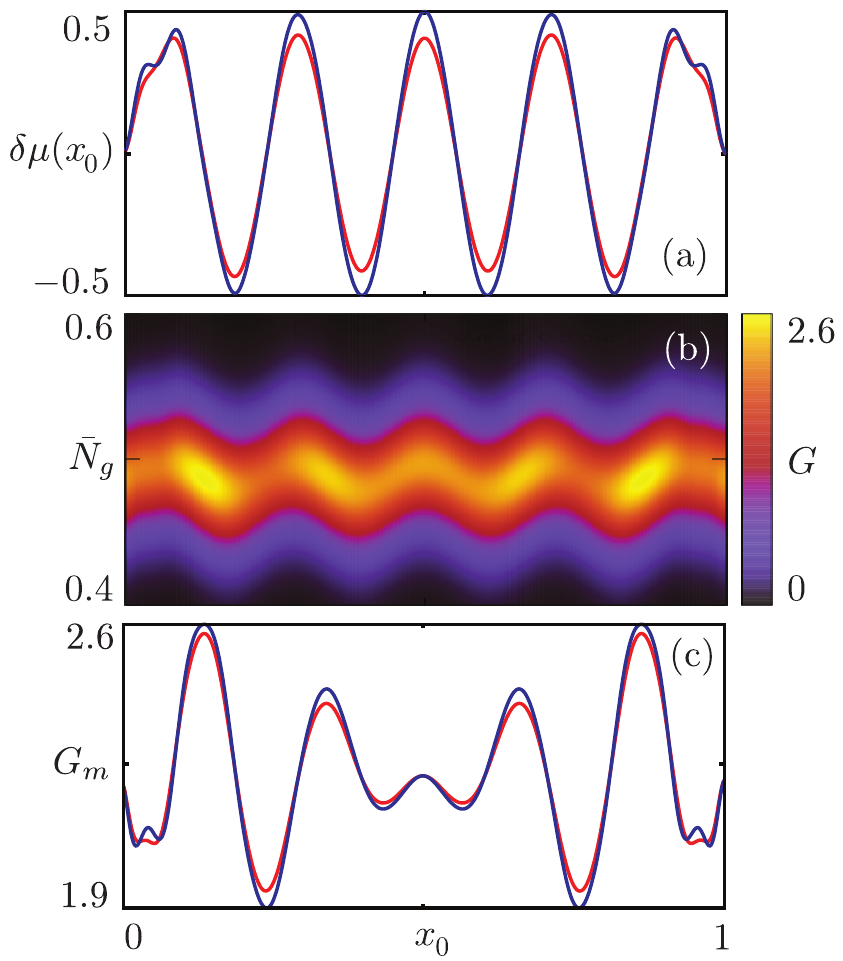}
  \caption{Transport properties of a suspended CNT dot near the transition between 16 and 17 electrons for mild interactions, $g=0.7$, when the AMF tip is placed at $x_{0}$. (a) Corrections to the chemical potential $\delta\mu(x_0)$ (units $V^{(F)}$) as a function of $x_0$ (units $L$); (b) Density plot of the linear conductance $G$ (units $e^2\gamma_{0}^{(S)}/\omega_{2}$) as a function of $x_{0}$ (units $L$) and $\bar{N}_g=N_{g}-3\omega_{2}/8E_{1}$; (c) Maximum of the linear conductance as a function of $x_{0}$ (units $L$). In panels (a) and (c), blue (red) curves have been calculated neglecting (including) electron-vibron coupling. Other parameters: $V^{(F)}=V^{(W)}=0.05\omega_{2}$, $k_{B}T=0.1\Omega_{0}$, $\lambda_{m}=2$, $v_{F}/v_{s}=4.6$ appropriate for a semiconducting CNT and $\tilde{a}=L/50$.}
  \label{fig:fig2}
  \end{center}
\end{figure}
Figure~\ref{fig:fig2} shows the calculated linear conductance and chemical potential shift for a CNT tuned about the transition between $N=16$ electrons ($N_0=16$, $N_1=0$) and $N=17$ electrons ($N_0=16$, $N_1=1$) for the case of mild interactions, $g=0.7$. Panel (a) shows the oscillations of the chemical potential shift $\delta\mu(x_{0})$ as a function of the tip position $x_{0}$. In the weak interaction regime, Friedel oscillations dominate over the Wigner ones: as a result a typical shape with $N_0/4+1=5$ maxima and $N_0/4=4$ minima is shown. This corresponds to the wavelength of the Friedel oscillations which is not determined by the total number of electrons but by the number of electrons in the sector $\alpha,\sigma$ involved in transport, in analogy with the behaviour of a two-channel quantum wire~\cite{sgm2,AFM}. The blue curve represents the results in the absence of electron-vibron coupling: only minor ripples near the dot edges signal the weak influence of the Wigner molecule. The red curve shows the results obtained including the electron-vibron coupling: a suppression of the oscillations is evident, which also washes away the small features near the border. The overall shape of $\delta\mu(x_{0})$ is, however, completely analogous to the static case. Here and in the following, we have chosen a semiconducting CNT with a fairly small Fermi velocity $v_{F}/v_{s}=4.6$, in order to enhance the electron-vibron coupling. Panel (b) shows the linear conductance as a color map, as a function of the tip position and the number of charges induced by the gate. The latter quantity is responsible for tuning $\mu_{0}$ and hence for bringing the dot into the resonance, which occurs when $\mu=\mu_{0}+\delta\mu(x_0)=(k_{B}T/2)\ln(4)$, see Eq.~(\ref{eq:cond}). Since $\delta\mu(x_0)$ oscillates, the value of $N_{g}$ needed to obtain the resonance fluctuates, allowing to map the shift in chemical potential. Also the intensity of the conductance oscillates as the tip is scanned along the CNT. Panel (c) shows the height of the maximum $G_{m}=\mathrm{max}_{N_{g}}\{G\}$ as a function of $x_{0}$. The overall qualitative features are analogous to those of the chemical potential with the presence of 5 peaks and 4 valleys. The static CNT displays small ripples at the border of the dot, similarly to the behaviour of the chemical potential. However, such features basically disappear as a vibrating dot is considered, along with a suppression of the conductance.
\begin{figure}
\begin{center}
\includegraphics[width=7cm,keepaspectratio]{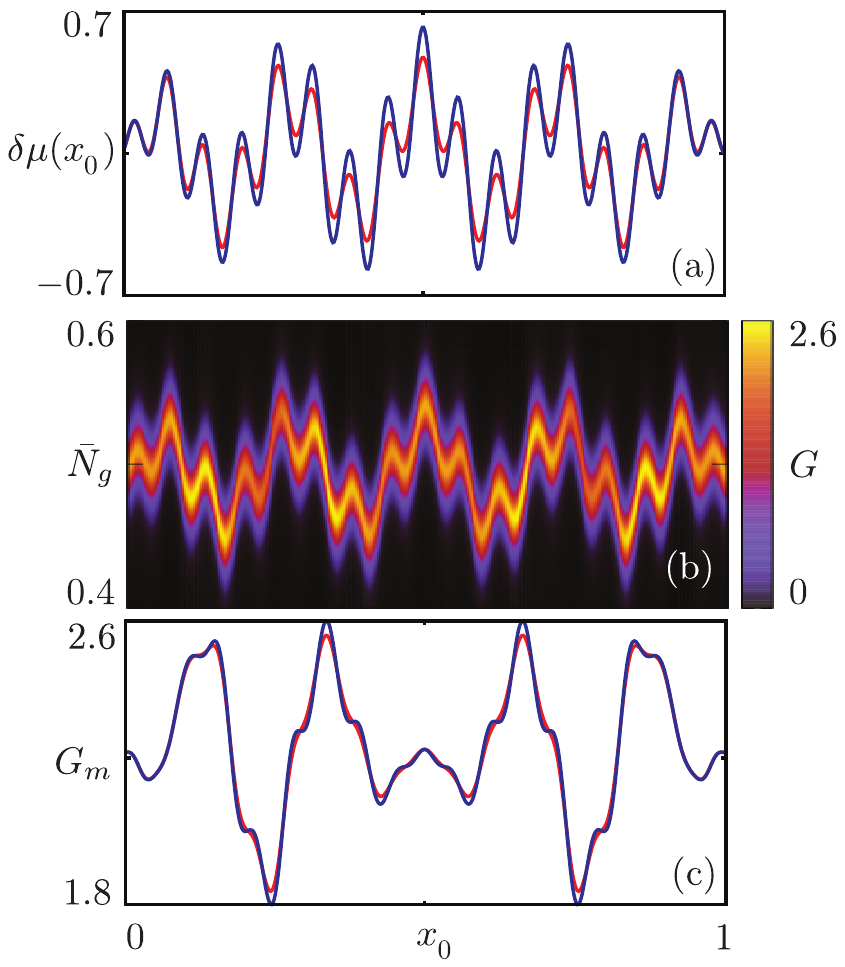}
\caption{Same as in Fig.~\ref{fig:fig2} but for a strongly interacting CNT, with $g=0.2$.}
 \label{fig:fig3}
  \end{center}
\end{figure}
Figure~\ref{fig:fig3} shows the case of a strongly interacting CNT, most favorable to observe the emergence of Wigner correlations. The chemical potential shift now distinctly shows 17 maxima and 16 minima, with an oscillation characterized by the Wigner wavelength. In the case of a vibrating CNT, a suppression of the oscillations is observed. Still, the qualitative shape of the chemical potential shift is unaffected. Indeed, as $g\to 0$, the parameter $r$ increases, which leads to $\alpha_{0}(x)\to 0$, thus making the contribution due to vibrons small. Also in this case the oscillations of the intensity of the conductance show oscillations with a period shorter than the Friedel one, However, the regime of a Wigner molecule is not fully developed in this quantity. The small suppression of the differential conductance maximum in the case of a vibrating dot does not alter qualitatively the results.\\
In conclusion we have calculated the correction to the chemical potential and the linear conductance of a 1D quantum dot built in an interacting and suspended CNT, perturbed by a charged AFM tip capacitively coupled to the dot. We demonstrated that both quantities depend on the electron-vibron coupling, whose effect is to suppress the oscillations of the chemical potential and of the conductance, induced by the Wigner and Friedel oscillations of the density. We have also shown that for realistic devices with a fairly strong electron-vibron coupling this suppression is not enough to produce qualitative modifications of the oscillating patterns. Indeed, in the case of strongly interacting nanotubes, the large mismatch of velocities between the plasma modes and the strain implies a strong reduction of the effects induced by vibrons, producing an even more favorable situation to observe the effects of Wigner molecules in a transport experiment.\\

\noindent {\it Acknowledgements.} Financial support by the EU- FP7 via ITN-2008-234970 NANOCTM is gratefully acknowledged.

\end{document}